\documentclass[twocolumn,epsfig,aps]{revtex4}
\def\>{\rangle}

\begin{document}
\draft
\title{A criterion for testing multi-particle NPT entanglement}
\author{B. Zeng$^1$, D. L. Zhou$^2$, P. Zhang$^3$, Z. Xu$^{1,2}$, and L. You$^{3,4}$}
\address{$^1$Department of Physics, Tsinghua University,
Beijing, 100084, China}
\address{$^2$Center for Advanced Study, Tsinghua University,
Beijing, 100084, China}
\address{$^3$Institute of Theoretical Physics, The Chinese
Academy of Sciences, Beijing 100080, China}
\address{$^4$School of Physics, Georgia Institute of Technology,
Atlanta, GA 30332, USA}
\date{\today}

\begin{abstract}
We revisit the criterion of multi-particle entanglement based on the overlaps
of a given quantum state $\rho$ with maximally entangled states. For a system
of $m$ particles, each with $N$ distinct states, we prove that
$\rho$ is $m$-particle negative partial transpose (NPT)
entangled, if there exists a maximally
entangled state $|{\rm MES}\rangle$, such that $\langle{\rm
MES}|\rho|{\rm MES}\rangle>{1}/{N}$. While this sufficiency condition
is weaker than the Peres-Horodecki criterion in all cases, it applies to
multi-particle systems, and becomes
especially useful when the number of particles ($m$) is large. We
also consider the converse of this criterion and illustrate its
invalidity with counter examples.
\end{abstract}

\maketitle
%\narrowtext

Quantum entanglement is among the most intriguing properties of
a composite quantum system. It usually refers to any
inseparable correlations that are stronger than those allowed
by classical physics.
In recent years, it has become increasingly clear that
{\it entanglement} is not only of interest to the understanding
of a proper interpretation of the foundations of quantum mechanics,
it also represents a useful resource for quantum information
processing and quantum computation \cite{Ni}.

Despite its importance, much remains to be learned about the
mathematical characterization of multi-particle entanglement. For
a $m$-particle pure state $|\psi\rangle$, it is well-known that
the necessary and sufficient condition for it to be $m$-particle
entangled, i.e. when $|\psi\rangle$ does not take the form of
$|\psi_{1}\rangle\otimes|\psi_{2}\rangle$, where
$|\psi_{1}\rangle$ is a $k$-particle state $(1\leq{k}\leq{m-1})$
and $|\psi_{2}\rangle$ is a state of the other $m-k$ particles, is
that all $1-{\rm tr}(\rho_{k}^{2})\neq 0$, where $\rho_{k}$ the
reduced density matrix (of state $|\psi\rangle$) for any
$k$-particles. In practice, this criterion for multi-particle
entanglement is difficult to apply, especially when $m$ is large
as the task of verification increases exponentially with $m$. The
case of mixed states is more difficult, even for bi-particle
systems \cite{Ho}. Instead of {\it entanglement}, inseparability
is often used for non-local quantum correlations of a mixed state $\rho$.

For bi-particle mixed states, it is well-known that the
Peres-Horodecki criterion is arguably the most useful sufficient
condition for quantum entanglement. It is also a necessary
condition for systems of reduced dimensions $2\otimes2$ and
$2\otimes3$ \cite{Pe,Ho1}. However for higher dimensional cases it
fails to be a neessary one since there exists a large number of
positive partial transpose (PPT) states \cite{Ho2,Be1}. Recently,
it has been successfully generalized to the special case of
continuous variables \cite{Du,Si}. Unfortunately, even to test NPT
entanglement, applying this criterion to a multi-particle system
meets the same computational challenge as in the pure state case.

Another useful concept of mixed state quantum correlation is {\it
distillability}, which measures the amount of pure state
entanglement `distillable' from a mixed state \cite{Le}.
Intuitively, one expects that if the overlap of a state $\rho$
with a maximally entangled state (MES) (sometimes also called
m-GHZ state \cite{GHZ}) is sufficiently high, then the state
$\rho$ must become distillable. These type of questions were first
considered by Bennett {\it et al.} \cite{Be2}. They introduced the
notion of a fully entangled fraction, or fidelity, to deal with
the relationship between entanglement of formation and
entanglement distillation. A necessary and sufficient condition
for a two qubit state $\rho$ to be distillable was later found to
be that the state $\rho$ is distillable if and only if its
fidelity is greater than ${1}/{2}$ \cite{Be3}. This result was
subsequently generalized to the case of higher dimensional
bi-particle states by Horodeckis and a sufficient condition for
N-dimensional bi-particle NPT entanglement is obtained
\cite{Ho,Ho3}. Conveniently stated, the result of Horodeckis
becomes: if the fidelity of $\rho$ is greater than ${1}/{N}$ then
the state of $\rho$ is NPT entangled.

In this paper we revisit a sufficient condition for $m$-particle
entanglement of a state $\rho$ based on its overlaps with
maximally entangled states. For a system of $m$ qubits, this
sufficiency condition first appeared in Ref. \cite{Sa}, and
was also discussed previously in Refs. \cite{Se,xs}.
Surprisingly, we find the Horodeckis' result
is valid for systems involving $m>2$ parties and the proof is
fairly simple. This naturally leads to the formulation of a new
multi-particle NPT entanglement criterion, which  is relatively easy
to apply, and thus seems particularly useful when $m$ is large.
Before we present a
proof for the new criterion, we first prove the following lemma, which
gives a simple criterion for testing multi-particle entanglement.

    \textbf{Lamma} For an arbitrary $N$-state $m$-particle
    separable state $\rho$, i.e. $\rho$
    can be written as $\sum_{i}p_{i}\rho_{1}^{(i)}\otimes\rho_{2}^{(i)}$
    (where $\rho_{1}^{(i)}$ is a $k$-particle state $(1\leq{k}\leq{m-1})$
    and $\rho_{2}^{(i)}$ is the state of the other $m-k$
    particles, $p_{i}>{0}$ and $\sum_{i}p_{i}=1$), then
    $\langle{\rm MES}|\rho|{\rm MES}\rangle\leq{1}/{N}$ for any MES.

    Proof: We first prove the pure state case, i.e. let $|\psi\rangle$ be a $m$-particle (each with $N$-state)
    pure state, if $|\psi\rangle$
    can be written as $|\psi_{1}\rangle\otimes|\psi_{2}\rangle$
    (where $|\psi_{1}\rangle$ is a $k$-particle state $(1\leq{k}\leq{m-1})$
    and $|\psi_{2}\rangle$ is the state of the other $m-k$
    particles), then
    $|\langle{\rm MES}|\psi\rangle|^{2}\leq{1}/{N}$ for any MES.

    We note that any $N$-state $m$-particle MES can be written as
    $|{\rm MES}\rangle=\sum_{i=1}^{N}|\!\underbrace{ii\cdots i}_{m}\rangle/{\sqrt{N}}$
    in a properly chosen basis. Let $a_{i}=\langle\underbrace{ii\cdots i}_{k}|\psi_{1}\rangle$
    and $b_{i}=\langle\underbrace{ii\cdots i}_{m-k}|\psi_{2}\rangle$,
    then we have $\sum_{i}|a_{i}|^{2}\leq{1}$,
    $\sum_{i}|b_{i}|^{2}\leq{1}$, and
    \begin{eqnarray}
    |\langle{\rm MES}|\psi\rangle|^{2}
    &=&\frac{1}{N}\left|\sum_{i=1}^{N}{a_{i}b_{i}}\right|^{2}
    \leq\frac{1}{N}\left(\sum_{i=1}^{N}|{a_{i}b_{i}}|\right)^{2}\nonumber\\
    &\leq&\frac{1}{N}\left(\sum_{i=1}^{N}\frac{|{a_{i}|^2+|b_{i}}|^2}{2}\right)^{2}
    \leq{\frac{1}{N}}.
    \label{eq1}
    \end{eqnarray}
    The equality $|\langle{\rm MES}|\psi\rangle|^{2}={1}/{N}$
    holds iff
    \begin{eqnarray}
    |a_{i}|&=&|b_{i}|,\nonumber\\
    \arg(a_{i}b_{i})&=&{\rm const.},\nonumber\\
    \sum_{i=1}^{N}|a_{i}|^{2}&=&\sum_{i=1}^{N}|b_{i}|^{2}=1.
    \label{eq2}
    \end{eqnarray}
    where $\arg(.)$ refers to the argument of the complex number variable.

    Give a mixed state $\rho=
    \sum_{i}p_{i}\rho_{1}^{(i)}\otimes\rho_{2}^{(i)}$,
    we can rewrite $\rho_{1}^{(i)}$ and $\rho_{2}^{(i)}$ in their diagonal
    representation, i.e.

    \begin{eqnarray}
    \rho&=&\sum_{i}p_{i}\left(\sum_{j}q_{j}|\psi_{1}^{(i)}\rangle_{j}{_{j}\!\langle\psi_{1}^{(i)}}|\right)
    \otimes\left(\sum_{l}r_{l}|\psi_{2}^{(i)}\rangle_{l}{_{l}\!\langle\psi_{2}^{(i)}}|\right)\nonumber\\
    &=&\sum_{i,j,l}p_{i}q_{j}r_{l}
    \left(|\psi_{1}^{(i)}\rangle_{j}\otimes|\psi_{2}^{(i)}\rangle_{l}{_{j}\!\langle\psi_{1}^{(i)}}|
    \otimes{_{l}\!\langle\psi_{2}^{(i)}|}\right),
    \end{eqnarray}
    where $\sum_{j}{q_j}=1$ and $\sum_{l}{r_l}=1$.
    Using the result of Eq. (\ref{eq1}) for a pure state, we get
    \begin{eqnarray}
    \langle{\rm MES}|\rho|{\rm MES}\rangle\leq\frac{1}{N}
    \sum_{i,j,l}p_{i}q_{j}r_{l}=\frac{1}{N}.
    \end{eqnarray}

This proof naturally leads to the following criterion for
multi-particle entanglement: for any state $\rho$ of a $N$-state
$m$-particle system, if there exists a maximally entangled state
$|{\rm MES}\rangle$, such that $\langle{\rm MES}|\rho|{\rm
MES}\rangle>{1}/{N}$, then $\rho$ is $m$-particle entangled.
This generalizes a similar criterion for the case
of many qubits discussed before in Refs. \cite{Sa,Se,xs}

The main result of this paper is to prove that the above criterion
is a criterion for multi-particle NPT entanglement, i.e.

\textbf{Theorem (The MES criterion)} For any state $\rho$ of a
$N$-state $m$-particle system, if there exists a maximally
entangled state $|{\rm MES}\rangle$, such that $\langle{\rm
MES}|\rho|{\rm MES}\rangle>{1}/{N}$, then $\rho$ is $m$-particle
NPT entangled, i. e. $\rho ^{T_B}$ is no longer positive for
partial transpose of any $k$-particle subsystems, where
$(1\leq{k}\leq{m-1})$.

Proof: We prove the negative inverse proposition of the criterion,
i. e. for any state $\rho$ of a $N$-state $m$-particle system, if
there is a partial transpose operator $T_B$ acting on any of its
$k$-particle subsystem, where $(1\leq{k}\leq{m-1})$, s.t. $\rho
^{T_B}\geq 0$, then $|\langle{\rm
MES}|\psi\rangle|^{2}\leq{1}/{N}$ for any MES.

    To prove this, we first write an arbitrary MES as
    $|{\rm MES}\rangle=\sum_{i=1}^{N}|\!\underbrace{ii\cdots i}_{m}\rangle/{\sqrt{N}}$
    in a properly chosen basis. Let $P_{+}=|{\rm MES}\rangle\langle {\rm
    MES}|$, i. e. a projection operator onto this multi-particle maximally
    entangled state. For a given partial transpose on any $k$-particle
    subsystem, we define a map
    \begin{eqnarray}
    |\underbrace{ii\cdots i}_{k}\rangle\longrightarrow |i\rangle_A,\nonumber\\
    |\underbrace{ii\cdots i}_{m-k}\rangle\longrightarrow
    |i\rangle_B,
    \end{eqnarray}
    general states in the two subspaces
    \begin{eqnarray}
    |\psi_\alpha\rangle_A=\sum\limits_{i=1}^{N} \alpha_i|i\rangle_A,\nonumber\\
    |\psi_\beta\rangle_B=\sum\limits_{i=1}^{N}\beta_i |i\rangle_B,
    \end{eqnarray}
    and a flip operator $V=\sum_{ij}\left| i\right\rangle _{AA}\left\langle j\right| \otimes \left|
j\right\rangle _{BB}\left\langle i\right|$, such that
    \begin{eqnarray}
    V|\psi_\alpha\rangle_A\otimes|\psi_\beta\rangle_B=|\psi_\beta\rangle_A\otimes
    |\psi_\alpha\rangle_B.
    \end{eqnarray}

    It is then easy to show that
    \begin{eqnarray}
    P_{+}^{T_B}=\frac{1}{N}V,
    \end{eqnarray}
    which leads to
    \begin{eqnarray}
    {\rm tr}(\rho P_{+})={\rm tr}(\rho^{T_B}P_{+}^{T_B})=\frac{1}{N}{\rm tr}(\rho^{T_B}V).
    \end{eqnarray}

    Since $\rho$ is PPT under $T_{B}$, $\rho^{T_B}$ is a
    legitimate state and consequently ${\rm tr}(\rho^{T_B}V)\leq 1$, we thus conclude
    \begin{eqnarray}
    \langle{\rm MES}|\rho|{\rm MES}\rangle\leq\frac{1}{N}.
    \end{eqnarray}

    This completes the proof of the criterion, which is indeed a
    multi-particle generalization of the bipartite conclusion in
    Ref. \cite{Ho}.

This MES criterion is very useful when $m$ is large, e.g. in the
case of massive entanglement of Bose condensed atoms
\cite{You,So}. Now consider the simple example of an arbitrary
pure state of a $N$-state $m$-particle system,
$|\psi\rangle=\sum_{i_{1},i_{2}\cdots ,i_{m}} a_{i_{1},i_{2}\cdots
,i_{m}}|i_{1}i_{2}\cdots i_{m}\rangle$, we can immediately state
that it is $m$-particle entangled if $|\sum_{i}a_{i,i\cdots
,i}|>1$.

As an example, we consider the two qubit state
\begin{eqnarray}
\rho &=&p |\psi_1\rangle\!\langle\psi_1|+
     (1-p) |\psi_2\rangle\!\langle\psi_2|, \nonumber\\
|\psi_1\rangle &=&  \frac{1}{\sqrt{2}}(|00\rangle-|11\rangle), \nonumber\\
|\psi_2\rangle &=&  |00\rangle, \label{rho}
\end{eqnarray}
as considered by Horodeckis \cite{Ho1}, where $0\leq p\leq 1$.
Making use of the fact that
$\langle\Psi^{-}|\rho|\Psi^{-}\rangle=(1+p)/{2}$ and
$|\Psi^{-}\rangle=(|00\rangle-|11\rangle)/{\sqrt{2}}$, we
immediately find that $\rho$ is entangled for any $p\neq0$, in
agreement with the result of Ref. \cite{Ho1}.

It is also straightforward to show that the MES criterion gives
the upper bound for entanglement as found in NMR based systems \cite{NMR}.
Let's consider the bi-particle state of spin-$({N-1})/{2}$ particles
\begin{eqnarray}
\rho=(1-\epsilon)\frac{I}{N^2}+\epsilon{|{\rm MES}\rangle}\!{\langle {\rm MES}|},
\end{eqnarray}
with $I$ the identity operator. Since $\langle {\rm MES}|I|{\rm
MES}\rangle=1$, we find the state of $\rho$ is NPT entangled if
\begin{eqnarray}
{\langle {\rm MES}|\rho|{\rm MES} \rangle}
=\frac{(1-\epsilon)}{N^2}+\epsilon>\frac{1}{N},
\end{eqnarray}
i.e. $\epsilon>{1}/({N+1})$ \cite{NMR}. According
to arguments as given Ref. \cite{NMR}, this is also true
for systems with $m>2$ particles. It thus gives
the same upper bound for entanglement of NMR based systems.

It is not difficult to show that our criterion can be generalized to
the case when each different particle may have diffident number of states
$\{N_{j}\}_{j=1}^{m}$. All it takes is to replace the $N$ by
$\min(N_{1},N_{2}\cdots ,N_{m})$, for the
definition of MES in such a case.

We now consider the converse of our criterion. To answer this
question, we first focus on the case of pure quantum states. It is
also noticed that for pure states, any entangled state is NPT. For
a $2$-particle pure state system, the answer is an affirmative yes
because of the following corollary.

\textbf{Corollary 1} An arbitrary $N$-state $2$-particle pure
state $|\psi\rangle$ is entangled if and only if there exists a
MES, such that $|\langle{\rm MES}|\psi\rangle|^{2}>{1}/{N}$.

    Proof: The ``if" part comes from our MES criterion.
    The ``only if" part is proved using the well
    known result that any bi-particle pure state can be written in the form
    of a Schmidt decomposition \cite{Sc}
    \begin{eqnarray}
    |\psi\rangle=\sum_{i=1}^{n}k_{i}|ii\rangle,
    \end{eqnarray}
    with $k_{i}>{0}$ and $\sum_{i=1}^{n}k_{i}^2=1$.
$n$ ($1\leq{n}\leq{N}$) is the Schmidt number of state $|\psi\rangle$,
which is entangled iff $n>1$.
Now take $|{\rm MES}\rangle=\sum_{i=1}^{N}|ii\rangle/{\sqrt{N}}$,
for an entangled state $|\psi\rangle$ ($n>1$ and $k_i>0$)
    \begin{eqnarray}
    |\langle{\rm MES}|\psi\rangle|^{2}
    =\frac{1}{N}\left(\sum_{i=1}^n k_{i}\right)^2>
    \frac{1}{N}\sum_{i=1}^n k_{i}^2=\frac{1}{N},
    \end{eqnarray}
which completes our proof.

From this corollary we have the following corollary immediately.

    \textbf{Corollary 2} For any pure state $|\psi\rangle$ of
    a $N$-state $2$-particle system, if there exists a
    MES, such that $|\langle{\rm MES}|\psi\rangle|^{2}>{p}/{N}$, where
    $p$ is a natural number, then its Schmidt number $n$ must satisfy $n>p$.

    For $3$-particle systems, the following corollary 3 applies.

    \textbf{Corollary 3} If an arbitrary $N$-state $3$-particle
    pure state $|\psi\rangle$ is not $3$-particle entangled, then there exists a
    MES, such that $|\langle{\rm MES}|\psi\rangle|^{2}={1}/{N}$.

    Proof: We directly construct a MES to satisfy
    $|\langle{\rm MES}|\psi\rangle|^{2}={1}/{N}$.
    If $|\psi\rangle$ is not $3$-particle entangled, it has
    only two possible types of separability. Either
    it can be written as $|111\rangle$ in some particular basis, where
    apparently $|{\rm MES}\rangle=\sum_{i=1}^{N}|iii\rangle/{\sqrt{N}}$
    will meet our requirement; or one of the particles
    is entangled with a second particle, but the two entangled ones
    are not entangled with the third particle. Take
    $|\psi\rangle=|\psi\rangle_{1}\otimes|\psi\rangle_{23}$, for
    example, where $1$ and $23$ stand for particle indices $1$, $2$, and $3$.
    Considering the condition of Eq. (\ref{eq2}), we first write
    particle $2$ and $3$ in their Schmidt basis and then construct
    the basis where the state of particle $1$ can be expressed as
    $\sum_{i=1}^{N}k_{i}|i\rangle$,
    where $k_{i}\geq{0}$ are the Schmidt coefficients of state $|\psi\rangle_{23}$
    and $\sum_{i=1}^{N}k_{i}^2=1$. Thus  $|\psi\rangle$ can be rewritten as

    \begin{eqnarray}
    |\psi\rangle=\sum_{i=1}^{N}k_{i}|i\rangle\otimes\sum_{i=1}^{N}k_{i}|ii\rangle.
    \end{eqnarray}
It is then easy to check
that $|{\rm MES}\rangle=\sum_{i=1}^{N}|iii\rangle/{\sqrt{N}}$
again fulfills the requirement.

Because Schmidt decomposition generally does not apply to three
particle states, it is difficult to know whether the
converse proposition of our MES criterion is true for $m=3$. Even
in situations where the generalized Schmidt decomposition applies
\cite{Ca,Ac}, we have failed to prove the validity of the converse
analytically, although our numerical results indicate it to be
always true. This can perhaps be understood since
three-particle entanglement is
essentially different from that of two parties.

For four-qubit systems, we find that even the counterpart of
Corollary 3 does not exist. Take the state
$|\psi\rangle=|11\rangle\otimes(|11\rangle+|22\rangle)/{\sqrt{2}}$,
    for example, it is easy to show that this state can never satisfy
    the condition of Eq. (\ref{eq2}) in any basis,
i.e. $|\langle{\rm MES}|\psi\rangle|^{2}={1}/{2}$ can not be
    achieved. Our numerical results give
    $\max(|\langle{\rm MES}|\psi\rangle|^{2})={1}/{4}$, where the
    maximization is over all MES. Interestingly, our numerical
    simulations also show that Corollary 1 no longer holds for 4-particle systems.
    For example, the 4-particle $W$ state
    $(|0001\rangle+|0010\rangle+|0100\rangle+|1000\rangle)/{2}$ \cite{Dur}
    can only have a maximal projection onto MES of $0.347$.
    This result illustrates the essential difference between
    the four-particle entanglement and that of three parties.

These results and examples thus constitute a negative answer to the
converse proposition of our MES criterion.

For mixed states, our numerical results show that the
converse of the MES sufficiency criterion fails
even for two qubit systems.
Take the following state,
\begin{eqnarray}
\rho &=&p |\psi_1\rangle\!\langle\psi_1|+
     (1-p) |\psi_2\rangle\!\langle\psi_2|, \nonumber\\
|\psi_1\rangle &=&  a|00\rangle+b|11\rangle, \nonumber\\
|\psi_2\rangle &=&  a|01\rangle+b|10\rangle,
\label{rho}
\end{eqnarray}
with $a,b>0$ and $a^2+b^2=1$, for example, as considered by
Horodeckis in Ref. \cite{Ho1}. It is known to be entangled if and
only if $ab\neq 0$ and $p\neq{1}/{2}$. We find numerically,
however, that for many entangled states of this type,
$\max(\langle{\rm MES}|\rho|{\rm MES}\rangle)<{1}/{2}$. For
instance, when $a=0.6$, $b=0.8$, and $p=0.495$, state $\rho$ of
Eq. (\ref{rho}) is entangled, yet it is easy to check that
$\max(\langle {\rm MES}|\rho|{\rm MES}\rangle)=0.4949<{1}/{2}$.

In summary, we have proposed and proved a criterion, a sufficient condition
for testing multi-particle NPT entanglement, based on the maximal
overlap of a given quantum state with MES. We have shown that for
bi-particle pure state systems, this criterion is also a necessary
condition. In general, however, the converse proposition of our
criterion is not true, i.e. this MES
    criterion is not a necessary condition for multi-particle NPT
    entanglement. We believe our result will shed new light on
    the studies of multi-particle entanglement, and it is especially
    useful in practical problems
    where the particle number $m$ is large. We also hope that
    our result will help further investigations of
    multi-particle quantum state distillation.

We thank Mr. J. G. Xiang, Prof. X. F. Liu,
and Prof. C. P. Sun for helpful discussions. The work of D. L.
Zhou is partially supported by the National Science Foundation of China
(CNSF) grant No. 10205022. P. Zhang is supported by CNSF,
the Knowledged Innovation Program (KIP) of the Chinese
Academy of Science, and the National Fundamental Research Program
of China (No. 001GB309310). L. You acknowledges support from
NSF and CNSF(B). The work of Z. Xu is partially
supported by the CNSF grant No. 90103004 and 10247002.

\end{document}